\documentclass[sort]{svproc}
\usepackage{url}
\usepackage{todonotes}

\usepackage[utf8]{inputenc}
\usepackage{amsfonts}
\usepackage{amssymb}
\usepackage{enumerate}
\usepackage{lineno}
\usepackage{enumitem}
\usepackage{mathrsfs}
\usepackage{mathtools}

\usepackage{color}

\begin{document}
\mainmatter              
\title{An Algorithmic Information Distortion in Multidimensional Networks}
\titlerunning{Algorithmic Information Distortion}  
%
\toctitle{An Algorithmic Information Distortion in Multidimensional Networks}
\author{Felipe S. Abrah\~{a}o\inst{1} \and Klaus Wehmuth\inst{1} \and Hector Zenil\inst{2} \and Artur Ziviani\inst{1} }
\authorrunning{Felipe S. Abrah\~{a}o et al.} 
%
\tocauthor{Felipe S. Abrah\~{a}o, Klaus Wehmuth, Hector Zenil, Artur Ziviani}
\institute{National Laboratory for Scientific Computing (LNCC), 25651-075, Petropolis, RJ, Brazil.\\
	\email{fsa@lncc.br},\quad \email{klaus@lncc.br}, \quad \email{ziviani@lncc.br}\\
	\and
	Oxford Immune Algorithmics, RG1 3EU, Reading, U.K.\\ Algorithmic Dynamics Lab, Unit of Computational Medicine, Department of Medicine Solna, Center for Molecular Medicine, Karolinska Institute, SE-171 77, Stockholm, Sweden.\\ Algorithmic Nature Group, Laboratoire de Recherche Scientifique (LABORES) for the Natural and Digital Sciences, 75005, Paris, France.\\
	\email{hector.zenil@ki.se}}

\maketitle              

\begin{abstract}
	Network complexity, network information content analysis, and lossless compressibility of graph representations have been played an important role in network analysis and network modeling.
	As multidimensional networks, such as time-varying, multilayer, or dynamic multilayer networks, gain more relevancy in network science, it becomes crucial to investigate in which situations universal algorithmic methods based on algorithmic information theory applied to graphs cannot be straightforwardly imported into the multidimensional case.
	In this direction, as a worst-case scenario of lossless compressibility distortion that increases linearly with the number of distinct dimensions, this article presents a counter-intuitive phenomenon that occurs when dealing with networks within non-uniform and sufficiently large multidimensional spaces. 
	In particular, we demonstrate that the algorithmic information necessary to encode multidimensional networks that are isomorphic to logarithmically compressible monoplex networks may display exponentially larger distortions in the general case. 
\end{abstract}
%
\section{Introduction}\label{sectionIntroduction}

Algorithmic information theory (AIT) gives a set of formal universal tools for studying network complexity in the form of data compression, irreducible information content, or randomness of individual networks
\cite{Morzy2017a,Mowshowitz2012,Santoro2020nat,Zenil2018a}, 
especially in the case these networks were not generated, constructed, or defined by stochastic processes.
In addition, such an algorithmic approach to the study of complex networks (and not only graphs or networks, but tensors in general) has presented important refinements of more traditional statistical approaches, for example in the context of: automorphism group size \cite{Zenil2014}; graph summarization \cite{Zenil2019c,Zenil2016b}; typicality and null models by replacing the principle of maximum entropy with the principle of maximum algorithmic randomness \cite{Zenil2019b}; and the reducibility problem of multiplex networks into aggregate monoplex networks  \cite{Santoro2020nat}.
Since proper representations of multidimensional networks into new extensions of graph-theoretical abstractions have been one of the central topics of investigation in network science \cite{Boccaletti2014a,Domenico2013,Kivela2014,Lambiotte2019},
the situations in which previous methods based on AIT cannot be straightforwardly imported into the multidimensional case also become an important question. 

In this sense, we present in this article a theoretical analysis of worst-case distortions with respect to the algorithmic complexity of node-aligned multidimensional networks, in particular those represented by multiaspect graphs \cite{Wehmuth2016b,Wehmuth2017} with a large number of non-uniform aspects.

We show that in the general case of a multidimensional network there are algorithmic information distortions that grow linearly with the number of aspects and exponentially with respect to the algorithmic information of a monoplex network, whereas both the multidimensional network and this monoplex network are isomorphic structures.
The results in this article hold independently of the choice of the encoding method or the universal programming language.
This is because, given any two distinct encoding methods or any two distinct universal programming languages, the algorithmic complexity of an object represented in one way or the other can only differ by a constant whose value only depends on the choice of encoding methods or universal programming languages, but not on the choice of the object \cite{Calude2002,Downey2010,Li1997}.
That is, algorithmic complexity is pairwise invariant for any two arbitrarily chosen encodings.
Although only dealing with pairs of isomorphic objects in addition to this encoding invariance, we will see later on in Corollary~\ref{corResponseletterreply2} that algorithmic information distortions can in fact result from changing the multidimensional spaces into which isomorphic copies of the objects are embedded.
Thus, contributing to multidimensional network complexity analysis, our results establish a worst-case error margin for topological information content evaluation and lossless compressibility.
In addition, it shows the importance of multidimensional network encodings into which the multidimensional space itself is also encoded.

The article is organized as follows.
In Section~\ref{sectionGeneralBackground}, we recover necessary concepts, definitions, and results from the literature.
In Section~\ref{sectionRecursivelylabeledMAGs}, we study basic properties of encoded multiaspect graphs.
In Section~\ref{sectionResults}, we demonstrate the main results in Theorem~\ref{thmMAGisomorphism} and Corollaries~\ref{corResponseletterreply1} and~\ref{corResponseletterreply2}.  Section~\ref{sec:conc} concludes the paper. 

\section{Background}\label{sectionGeneralBackground}


\subsection{Multiaspect graphs }\label{subsubsectionGraphsandMAGs}

We directly base our notation regarding classical graphs on \cite{Bollobas1998,Brandes2005a,Diestel2017} and regarding multiaspect graphs (MAGs) on \cite{Wehmuth2016b,Wehmuth2017}.
In order to avoid ambiguities, minor differences in the notation from \cite{Wehmuth2016b,Wehmuth2017} will be introduced here.
In particular, the notation of 
MAG $ H = ( A , E ) $ is replaced with $ \mathscr{G} = \left( \mathscr{A} , \mathscr{E} \right) $, where 
the list $A$ of aspects is replaced with $\mathscr{A}$
and the composite edge set $ E $ is replaced with $ \mathscr{E} $.
This way, note that $\mathscr{A} =  \left( \mathscr{A}( \mathscr{G} )[1] , \dots , \mathscr{A}( \mathscr{G} )[i] , \dots , \mathscr{A}( \mathscr{G} )[p] \right) $ is a list of sets,
where each set in this list is an \emph{aspect} (or node dimension \cite{Abrahao2018enat}) denoted by $ \mathscr{A}( \mathscr{G} )[i] $.
The \emph{companion tuple} of a MAG $ \mathscr{G} $ becomes then denoted by $ \tau( \mathscr{G} ) $, where
\[
\tau( \mathscr{G} ) = \left( | \mathscr{A}( \mathscr{G} )[1] |, \dots , | \mathscr{A}( \mathscr{G} )[p] | \right)
\]
\noindent and $p$ is called the \emph{order} of the MAG.
As established in \cite{Wehmuth2017}, it is important to note that the companion tuple completely determines the size of the node-aligned set $
\mathbb{V}( \mathscr{G} ) = \bigtimes_{i=1}^{p} \mathscr{A}( \mathscr{G} )[i] 
$ of all \emph{composite vertices} $ \mathbf{v} = ( a_1 , \dots , a_p ) $ of $ \mathscr{G} $, and as a direct consequence also determines the size of the set 
$
\mathbb{E}(\mathscr{G}) = \mathbb{V}( \mathscr{G} ) \bigtimes \mathbb{V}( \mathscr{G} )
$
\noindent of all possible \emph{composite edges} $ e  = ( ( a_1 , \dots , a_p ) , ( b_1 , \dots , b_p ) )$ of $ \mathscr{G} $.
This way, for every MAG $ \mathscr{G} $, one has $ \mathscr{E}\left( \mathscr{G} \right) \subseteq \mathbb{E}(\mathscr{G}) $.
In this article, we employ hereafter the term \emph{multidimensional networks} to refer to node-aligned multidimensional networks that can be mathematically represented by MAGs.

In addition, we denote an \emph{undirected} MAG without \emph{self-loops} by $ \mathscr{G}_c = (\mathscr{A},\mathscr{E}) $, so that the set $ \mathbb{E}_c $ of all possible undirected and non-self-loop composite edges is defined by
\[
\mathbb{E}_c( \mathscr{G}_c ) \coloneqq   \{ \{ \mathbf{u} , \mathbf{v} \} \mid \mathbf{u},\mathbf{v} \in \mathbb{V}( \mathscr{G}_c )  \} 
\]
\noindent and $ \mathscr{E}\left( \mathscr{G}_c \right) \subseteq \mathbb{E}_c( \mathscr{G}_c ) $ always holds.
In a direct analogy to simple graphs, we refer to these MAGs  $ \mathscr{G}_c $ as \emph{simple} MAGs.

Regarding graphs, we follow the common notation and nomenclature \cite{Diestel2017,Bollobas1998,Harary2018}: 
we denote a general (directed or undirected) \emph{graph}  by $ G = ( V , E ) $, where $V$ is the finite set of vertices  and $E \subseteq V \times V$;
if a graph only contains undirected edges and does not contain self-loops, then it is called a \emph{simple} graph.
A graph $G$ is (vertex-)labeled when the members of $V$ are distinguished from one another by labels such as $ v_1, v_2, \dots , v_{ \left| V \right| } $.
If a simple graph is labeled by natural numbers, i.e., $ V = \{ 1 , \dots , n \} $ with $ n \in \mathbb{N} $, then it is called a \emph{classical} graph.

For the present purposes of this article, all graphs $G$ will be classical graphs and all MAGs will be simple MAGs.

One may adopt the convention of calling the elements of the first aspect of a MAG as \emph{vertices}, i.e., $ \mathscr{A}( \mathscr{G} )[1] = V( \mathscr{G} ) $.
Thus, a classical graph $G$ is a labeled \emph{first-order} (i.e., $p=1$) simple MAG $ \mathscr{G}_c $ with $ V( G ) = \mathbb{V}( \mathscr{G}_c ) = \{ 1 , \dots , \left| \mathbb{V}( \mathscr{G}_c ) \right| \}$.
Note that the term `vertex' should not be confused with term `composite vertex',
since they refer to same entity only in the case of first-order MAGs.

As established in \cite{Wehmuth2016b}, one can define a \emph{MAG-graph isomorphism} analogously to the classical notion of graph isomorphism:
a MAG $ \mathscr{G} $  is isomorphic to a graph $ G $ \textit{iff} there is a bijective function $ f : \mathbb{V}( \mathscr{G} ) \to V( G ) $ such that 
\[ e \in \mathscr{E}( \mathscr{G} ) \iff ( f( \pi_o( e ) ) , f( \pi_d( e ) ) ) \in E( G ) \text{ ,}\]
where $ \pi_o $ is a function that returns the origin composite vertex of a composite edge and $ \pi_d $ is a function that returns the destination composite vertex of a composite edge.
In order to avoid ambiguities with the classical isomorphism in graphs, which is usually a vertex label transformation, we call: such an isomorphism between a MAG and graph from \cite{Wehmuth2016b} a \emph{MAG-graph isomorphism}; the usual isomorphism between graphs \cite{Bollobas1998,Diestel2017} as \emph{graph isomorphism}; and the isomorphism between two MAGs $ \mathscr{G} $ and $ \mathscr{G}' $ (i.e., $ ( \mathbf{ u } , \mathbf{ v } ) \in \mathscr{E}( \mathscr{G} ) $  \textit{iff}  $ ( f( \mathbf{ u } ) , f( \mathbf{ v } ) ) \in \mathscr{E}( \mathscr{G}' ) $ ) as \emph{MAG isomorphism}.

It is shown in \cite{Wehmuth2016b} that a MAG is isomorphically equivalent to a graph:
\begin{theorem}\label{thmMAGisomorphism}
	For every MAG $ \mathscr{G} $ of order $p>0$, where all aspects are non-empty sets, there is a unique (up to a \emph{graph isomorphism}) graph $ G_{ \mathscr{G}} = \left( V , E \right) $ that is \emph{MAG-graph-isomorphic} to $ \mathscr{G} $, where
	\[
	| V( G_{ \mathscr{G} } ) | =  \prod\limits_{ n = 1 }^{ p } | \mathscr{A}( \mathscr{G} )[ n ] | = \left|  \mathbb{V}( \mathscr{G} ) \right|
	\text{ .}
	\]
\end{theorem}
However, we shall show in this article that, although both a MAG and its isomorphic graph can be encoded and both represent the same abstract relational structure, they may diverge in terms of compressibility or algorithmic information content in the general case.

\subsection{Algorithmic information theory (AIT)}\label{subsubsectionTMandAIT}

In this section, we recover some basic notations and definitions from the literature regarding algorithmic information theory (aka Kolmogorov complexity theory or Solomonoff-Kolmogorov-Chaitin complexity theory).
For an introduction to these concepts and notation, see \cite{Calude2002,Chaitin2004,Downey2010,Li1997}.

First, regarding some basic notation, 
let $ l(x) $ denote the length of a string $ x \in  \{ 0 , 1 \}^* $.
Let $ (x)_2 $ denote the binary representation of the number $ x \in \mathbb{N} $. 
Let $ x \upharpoonright_{n} $ denote the ordered sequence of the first $n$ bits of the fractional part in the binary representation of $ x \in \mathbb{R} $. That is, $ x \upharpoonright_{n} = x_1 x_2 \dots x_n $, where $ ( x )_2 = y.x_1 x_2 \dots x_n x_{n+1} \dots $ with $ y \in \{ 0 , 1 \}^* $ and $  x_1 , x_2 , \dots , x_n \, \in \{ 0 , 1 \} $.
We denote the result of the computation of an arbitrary Turing machine $\mathcal{M}$ with input $x \in L$ by the partial computable function 
$ \mathcal{M} \colon L \to L $.
Let $ \mathbf{L'_U} $ denote a binary \emph{prefix-free} (or \emph{self-delimiting}) universal programming language for a prefix universal Turing machine $\mathbf{U}$.
As usual, let $ \left< \, \cdot \, , \, \cdot \, \right> $ denote an arbitrary computable bijective pairing function \cite{Downey2010,Li1997}, which can be recursively extended in order to encode any finite ordered $n$-tuple in the form $ \left< \cdot \, , \, \dots \, , \, \cdot   \right> $.
Let $ w^* $ denote the lexicographically first $ \mathrm{p} \in \mathbf{L'_U} $ such that $ l(\mathrm{p}) $ is minimum and $ \mathbf{U}(p) = w $.
The algorithmic information content of an object $w$ is given by the (unconditional) \emph{prefix} \emph{algorithmic complexity} (also known as K-complexity, prefix Kolmogorov complexity, self-delimited program-size complexity, or Solomonoff-Kolmogorov-Chaitin complexity for prefix universal Turing machines), denoted by $ \mathbf{K}( w ) $, which is the length of the shortest program $w^* \in \mathbf{L'_U} $ such that $ \mathbf{U}(w^*) = w $.
The \emph{conditional} prefix algorithmic complexity of a binary string $ y $ given a binary string $ x $, denoted by $ K( y \, | x ) $, is the length of the shortest program $w \in \mathbf{L'_U} $ such that $ \mathbf{U}( \left< x , w \right> ) = y $.

With respect to weak asymptotic dominance of function $f$ by a function $g$, 
we employ the usual $f(x)=\mathbf{O}( g(x) )$ for the big \textbf{O} notation when $f$ is asymptotically upper bounded by $g$; 
and with respect to strong asymptotic dominance by a function $g$, we employ the usual $f(x)=\mathbf{o}( g(x) )$ when $g$ dominates $f$.

\section{Basic properties of encoded multiaspect graphs}\label{sectionRecursivelylabeledMAGs}

In a general sense, a MAG $ \mathscr{G}_c $ is said to be \emph{encodable} (i.e., recursively labeled, or with a univocal computably ordered data representation) given $ \tau( \mathscr{G}_c ) $ \textit{iff} there is an algorithm that, given the companion tuple $ \tau( \mathscr{G}_c ) $ as input, computes a bijective ordering of composite edges $ e \in \mathbb{E}_c( \mathscr{G}_c ) $ from composite vertices $ \mathbf{v} \in \mathbb{V}( \mathscr{G}_c ) $. 
That is, if the companion tuple $ \tau( \mathscr{G}_c ) $ of the MAG $ \mathscr{G}_c $ is known, then one can computably retrieve the position of any composite edge $ e = \{ \mathbf{ u } , \mathbf{v} \} $ in the chosen data representation of $ \mathscr{G}_c $ from both composites vertices $ \mathbf{ u } $ and $ \mathbf{v} $, and vice-versa.\footnote{ An explicit formal definition of encodability (i.e., recursive labeling) given the companion tuple $ \tau( \mathscr{G}_c ) $ can be found for example in \cite{Abrahao2018dextendedarxiv2020reportnat}.}
This way, following the usual definition of encodings, a MAG is encodable given $ \tau( \mathscr{G}_c ) $ \textit{iff} there is a algorithm that, given $ \tau( \mathscr{G}_c ) $ as input, can univocally encode any possible $ \mathscr{E}\left( \mathscr{G}_c \right) $ that shares the same companion tuple.

As expected, MAGs that have every element of its aspects labeled as a natural number can always be encoded.
The proof of Lemma~\ref{lemmaLabeledMAG} follows directly from the definition of MAG and the recursive bijective nature of the pairing function.\footnote{ The reader can found a proof of Lemma~\ref{lemmaLabeledMAG} in \cite{Abrahao2018dextendedarxiv2020reportnat}.}
In other words, a MAG can always be encoded if the information necessary to determine the companion tuple $ \tau( \mathscr{G}_c ) $ is previously given.
\begin{lemma}\label{lemmaLabeledMAG}
	Any arbitrary simple MAG $ \mathscr{G}_c $  with $ \mathscr{A}(\mathscr{G}_c )[ i ] = \{ 1, \dots , \left| \mathscr{A}(\mathscr{G}_c )[ i ] \right| \} \subset \mathbb{N} $, where $ \left| \mathscr{A}( {\mathscr{G}_c} )[ i ] \right| \in \mathbb{N} $ and $ 1 \leq i \leq p = \left|\mathscr{A}( {\mathscr{G}_c} )  \right| \in \mathbb{N} $, is encodable given $ \tau( \mathscr{G}_c ) $.
	
\end{lemma}
Note that there is then an algorithm that, given a bit string $ x \in \{ 0 , 1 \}^* $ of length $ \left| \mathbb{E}_c( \mathscr{G}_c )   \right|  $ as input, computes a composite edge set $ \mathscr{E}( \mathscr{G}_c ) $ and there is another algorithm that, given the encoded composite edge set $ \mathscr{E}( \mathscr{G}_c ) $ as input, returns a string $x$.
Such strings univocally represent (up to a MAG isomorphism or up to a reordering of composite edges) the characteristic function (or indicator function) of pertinence in the set $ \mathscr{E}( \mathscr{G}_c ) $, and thus we call them as \emph{characteristic strings} of the MAG:
\begin{definition}\label{BdefCharacteristicstringofasimpleMAG}
	Let $ \left( e_1 , \dots , e_{ \left| \mathbb{E}_c( \mathscr{G}_c )   \right| } \right) $ be any arbitrary ordering of all possible composite edges of a  simple MAG $ \mathscr{G}_c $.
	We say that a string $ x \in \{ 0 , 1 \}^* $ with $ l( x ) = \left| \mathbb{E}_c( \mathscr{G}_c )   \right| $ is a \emph{characteristic string} of a simple MAG $ \mathscr{G}_c $
	\textit{iff}, for every $ e_j \in \mathbb{E}_c( \mathscr{G}_c ) $,
	\[
	e_j  \in \mathscr{E}( \mathscr{G}_c  ) \iff \text{ the $j$-th digit in $x$ is $1$}
	\text{ ,}
	\]
	\noindent  where $ 1 \leq j \leq l( x ) $. 
	
\end{definition}

In order to ensure uniqueness of representations (now only up to a MAG automorphism) from which the algorithmic complexity are calculated, 
one may also choose to encode a MAG into a string-based representation using the pairing function $  \left< \cdot, \cdot \right> $ and a fixed ordering/indexing of the composite edges:
\begin{definition}\label{BdefPlaincomplexityofedgesets}
	Let $ \left( e_1 , \dots , e_{ \left| \mathbb{E}_c( \mathscr{G}_c )   \right| } \right) $ be any arbitrary ordering of all possible composite edges of a  simple MAG $ \mathscr{G}_c $.
	Then, $\left< \mathscr{E}( \mathscr{G}_c ) \right>$ denotes the composite edge set string
	$
	\left<  \left< e_1, z_1 \right>, \dots , \left< e_n , z_n \right>  \right> 
	$
	\noindent such that 
	\[
	z_i = 1  \iff e_i \in \mathscr{E}( \mathscr{G}_c )
	\text{ ,}
	\]
	\noindent where  $ z_i \in \{ 0 , 1 \} $ with $ 1 \leq i \leq n= | \mathbb{E}_c( \mathscr{G}_c ) | $. 
\end{definition}

In the case of graphs (or monoplex networks), we remember that 
there is always a unified and decidable way to encode a sequence of all possible undirected edges given any unordered pair $ \{ x , y \} $ of natural numbers $ x , y \in \mathbb{N} $, for example by encoding characteristic strings or adjacency matrices of arbitrary finite size.
Thus, encoding classical graphs with characteristic strings or with composite edge set strings is Turing equivalent and, therefore, it is also equivalent in terms of algorithmic information.
This is indeed an underlying basic property previously explored, e.g., in \cite{Buhrman1999,Zenil2018a,Zenil2019c}.
Additionally, in the case of infinite graphs, it was shown in \cite{Khoussainov2014} that encoding with infinite characteristic strings may generate other counter-intuitive phenomena with respect to algorithmic randomness.
The present article only deals with finite MAGs and graphs and with infinite families of finite MAGs and graphs.
Unlike classical graphs, we shall see later on in Corollary~\ref{corResponseletterreply1} that the relationship between characteristic strings and composite edge set strings in the case of simple MAGs does not behave so well.


Nevertheless, if the ordering assumed in Definition~\ref{BdefCharacteristicstringofasimpleMAG} matches the same ordering in Definition \ref{BdefPlaincomplexityofedgesets}, we have in Lemma~\ref{lemmaBasicMAGandstrings} below that both the MAG and its respective characteristic string are indeed ``equivalent'' in terms of algorithmic information, but except for the minimum information necessary to encode the multidimensional space (e.g., the algorithmic information of the encoded companion tuple in the form $ \left< \tau( \mathscr{G}_c ) \right> = \left< | \mathscr{A}( \mathscr{G} )[1] |, \dots , | \mathscr{A}( \mathscr{G} )[p] | \right> $).
As expected, the proof follows directly from the fact that an ordering of composite edges is always embedded into the notion of encodability by composite edge set strings (a complete proof can be found in \cite{Abrahao2018dextendedarxiv2020reportnat}).
\begin{lemma}\label{lemmaBasicMAGandstrings}
	Let $ x \in \{ 0 , 1 \}^* $.
	Let $ \mathscr{G}_c  $ be an encodable MAG given $ \tau( \mathscr{G}_c ) $ 
	such that $ x $ is the respective characteristic string.
	Then,
	\begin{align}
		\label{lemmaBasicMAGandstrings1}  \mathbf{K}( \left< \mathscr{E}( \mathscr{G}_c  ) \right> \, | \, x  )   & \leq  \mathbf{K}( \left< \tau( \mathscr{G}_c ) \right> ) + \mathbf{O}(1)  \\
		\label{lemmaBasicMAGandstrings2} \mathbf{K}( x \, | \, \left< \mathscr{E}( \mathscr{G}_c  ) \right>  )   & \leq  \mathbf{K}( \left< \tau( \mathscr{G}_c ) \right> ) + \mathbf{O}(1) \\
		\label{lemmaBasicMAGandstrings3} \mathbf{K}( x ) & = \mathbf{K}( \left< \mathscr{E}( \mathscr{G}_c  ) \right> )  \pm \mathbf{O}\big( \mathbf{K}( \left< \tau( \mathscr{G}_c ) \right> ) \big). 
	\end{align}

\end{lemma}
Note that, since a graph is a MAG of order $1$, and in this case characteristic strings and composite edge set strings become Turing equivalent,
then Lemma~\ref{lemmaBasicMAGandstrings} can be improved in the case of graphs so that one can eliminate $ \mathbf{K}( \left< \tau( \mathscr{G}_c ) \right> ) $ in Equations~\eqref{lemmaBasicMAGandstrings1} and \eqref{lemmaBasicMAGandstrings2}.
In addition, one can replace $ \mathbf{O}\big( \mathbf{K}( \left< \tau( \mathscr{G}_c ) \right> ) \big)  $ in Equation~\eqref{lemmaBasicMAGandstrings3} with $ \mathbf{O}\left( 1 \right)  $.


\section{A worst-case algorithmic information distortion from increasing the number of aspects}\label{sectionResults}

Basically, Lemma~\ref{lemmaBasicMAGandstrings} assures that the information contained in a simple MAG $ \mathscr{G}_c $ and in the characteristic string are the same, except for the algorithmic information necessary to computably determine the companion tuple. 
Unfortunately, one can show in Theorem \ref{BthmMAGgivencompaniontuple} below that this information deficiency between the data representation of a MAG (in the form e.g. $ \left< \mathscr{E}( \mathscr{G}_c ) \right> $) and its characteristic string cannot be much more improved in general.
In other words, as we show\footnote{ A preliminary version of Theorem~\ref{BthmMAGgivencompaniontuple} can be also found in \cite{Abrahao2018dextendedarxiv2020reportnat}.} in Theorem~\ref{BthmMAGgivencompaniontuple}, there are \emph{worst-case} scenarios of multidimensional spaces in which the algorithmic information necessary for retrieving the encoded form of the MAG from its characteristic string is close (except for a logarithmic term) to the upper bound given by Equation~\ref{lemmaBasicMAGandstrings1} in Lemma~\ref{lemmaBasicMAGandstrings}.
This shows a fundamental difference between encoding MAGs with characteristic strings (or, equivalently, adjacency matrices \cite{Zenil2019c}) and encoding MAGs with composite edge set strings.
\begin{theorem}\label{BthmMAGgivencompaniontuple}
	There are arbitrarily large encodable simple MAGs $ \mathscr{G}_c $ given $ \tau( \mathscr{G}_c ) $
	such that 
	\[
	\mathbf{K}( \left< \tau( \mathscr{G}_c ) \right> ) + \mathbf{O}(1)
	\geq
	\mathbf{K}( \left< \mathscr{E}( \mathscr{G}_c  ) \right> \, | \, x  ) 
	\geq  
	\mathbf{K}( \left< \tau( \mathscr{G}_c ) \right> ) - \mathbf{O}\Big( \log_{2}\big(  \mathbf{K}( \left< \tau( \mathscr{G}_c ) \right> ) \big) \Big)
	\] 
	with 
	$ \mathbf{K}( \left< \mathscr{E}( \mathscr{G}_c  ) \right> ) \geq p - \mathbf{O}(1) $
	and
	$ \mathbf{K}( x ) = \mathbf{O}\left( \log_{2}\left(  p \right) \right) $, where
	$ x $ is the respective characteristic string and $p$ is the order of the MAG $ \mathscr{G}_c $.
	\begin{proof}
		The main idea of the proof is to define an arbitrary companion tuple such that the algorithmic complexity of the characteristic string is sufficiently small compared to the algorithmic complexity of the companion tuple, while we can prove that there is a computable procedure that always recovers the companion tuple from $ \left< \mathscr{E}( \mathscr{G}_c  ) \right> $.
		First, let $ \mathscr{G}_c  $ be any simple MAG with $ \tau( \mathscr{G}_c ) = \left( | \mathscr{A}( \mathscr{G}_c )[1] |, \dots , | \mathscr{A}( \mathscr{G}_c )[p] | \right) $ such that
		\begin{equation*}
			\begin{aligned}
				\mathscr{A}( \mathscr{G}_c )[i] = \left\{ 1 , 2 \right\} &\iff \text{the $i$-th digit of $w$ is $1$} \\
				\mathscr{A}( \mathscr{G}_c )[i] = \left\{ 1 \right\} &\iff \text{the $i$-th digit of $w$ is $0$} 
			\end{aligned}
		\end{equation*}
		where $ p \in \mathbb{N} $ and $ w \in \{ 0 , 1 \}^* $ are arbitrary. 
		Since $w$ is arbitrary, let $w$ be a long enough finite initial segment of a $1$-random real number $y$.
		Remember that, if $y$ is a \emph{$1$-random real number} (i.e., an algorithmically random infinite sequence \cite{Calude2002,Downey2010}), then $ \mathbf{K}( y \upharpoonright_n ) \geq n - \mathbf{O}(1) \text{ ,} $ \noindent where $ n \in \mathbb{N} $ is arbitrary.
		From Lemma~\ref{lemmaLabeledMAG}, we have that $ \mathscr{G}_c  $  is encodable given $ \tau( \mathscr{G}_c ) $.
		Therefore, there is a program $\mathrm{q}$ that represents an algorithm running on a prefix universal Turing machine $ \mathbf{U} $ that proceeds as follows:
		
		\begin{enumerate}[label=(\roman*),topsep=0.8pt,leftmargin=3\parindent]
			\item receive  $ \left< \mathscr{E}( \mathscr{G}_c  ) \right>^{*}  $ as input; 
			\item calculate the value of $ \mathbf{U}\left( \left< \mathscr{E}( \mathscr{G}_c  ) \right>^{*} \right) $ and build a sequence $ \left<  e_1, \dots , e_n \right> $ of the composite edges $ e_i \in \mathbb{E}_c( \mathscr{G}_c  )$ in the exact same order that they appear in $ \left< \mathscr{E}( \mathscr{G}_c  ) \right>  = \mathbf{U}\left( \left< \mathscr{E}( \mathscr{G}_c  ) \right>^{*} \right) $; 
			\item build a finite ordered set 
			\[
			\mathbb{V}' \coloneqq \left\{ \mathbf{ v } 
			\middle\vert
			e'
			\in \left<  e_1, \dots , e_n \right> 
			\text{ , where }
			\left( e' = \left< \mathbf{ v } , \mathbf{ u } \right> 
			\lor
			e' = \left< \mathbf{ u } , \mathbf{ v } \right> \right)
			\right\}
			\text{ ;}
			\]
			\item build a finite list $ \left[ A_1 , \dots , A_p  \right] $ of finite ordered sets
			\[
			A_i \coloneqq
			\left\{
			a_i
			\middle\vert
			a_i
			\text{ is the $i$-th element of }
			\mathbf{ v } = \left( a_1, \dots, a_p \right) \in \mathbb{V}' 
			\right\}
			\text{ ,}
			\] 
			where $p$ is finite and is smaller than or equal to the length of the longest $ \mathbf{ v } \in \mathbb{V}'  $;
			\item for every $i$ with $ 1 \leq i \leq p $, make
			\[
			z_i \coloneqq \left| A_i \right|
			\text{ ;}
			\]
			\item return the binary sequence $ s = x_1 x_2 \cdots x_p  $ from
			\begin{equation*}
				\begin{aligned}
					z_j \geq 2&\iff x_j = 1  \\
					z_j = 1 &\iff x_j = 0 
					\text{ .}
				\end{aligned}
			\end{equation*}
		\end{enumerate}
		Therefore, from our construction of $ \mathscr{G}_c   $, we will have that 
		\begin{equation}\label{equationKrandomrealandcompositeedgesetstring}
			\mathbf{K}(  y \upharpoonright_{ p }) \leq l\left( \left< \left< \mathscr{E}( \mathscr{G}_c  ) \right>^* ,  \mathrm{q} \right> \right)\leq \mathbf{K}( \left< \mathscr{E}( \mathscr{G}_c  ) \right>  ) + \mathbf{O}(1)
		\end{equation}
		holds by the minimality of $ \mathbf{K}( \cdot ) $ and by our construction of $\mathrm{q}$.
		Moreover, one can trivially construct an algorithm that returns $ y \upharpoonright_{ p } $ from the chosen companion tuple $ \tau( \mathscr{G}_c ) $ and another algorithm that performs the inverse computation. 
		This way, we will have that
		\begin{equation}\label{equationCompaniontuplefromprefixesofrandomreals}
			\mathbf{K}( \left< \tau( \mathscr{G}_c ) \right> )
			\leq 
			\mathbf{K}( y \upharpoonright_{ p } ) + \mathbf{O}(1)
			\leq
			p + \mathbf{O}\left( \log_{2}\left( p \right) \right)
		\end{equation}
		and, since $y$ is $1$-random,
		\begin{equation}\label{equationPrefixesofrandomrealsfromcompaniontuple}
			p - \mathbf{O}(1)
			\leq
			\mathbf{K}( y \upharpoonright_{ p } )
			\leq 
			\mathbf{K}( \left< \tau( \mathscr{G}_c ) \right> ) + \mathbf{O}(1).
		\end{equation}
		Additionally, since $ \mathscr{E}( \mathscr{G}_c  )    $ and $p$ were arbitrary, we can choose any characteristic string $ x $ such that
		\begin{equation}\label{equationLogarithmiccompressiblechasracteristicstring1}
			\mathbf{K}( x ) 
			= \mathbf{O}\left( \log_{2}\left(  p \right) \right)
		\end{equation}
		holds.
		For example,\footnote{This is only an example. In fact, one can choose any characteristic string $x$ in which $ \mathbf{K}(x) \leq \mathbf{O}\left( \log_{2}( p ) \right) $ always holds.} one can take a trivial $ x $ as a binary sequence starting with $1$ and repeating $0$'s until the length matches the total number of all possible composite edges
		\begin{equation}\label{equationBorelnormalityindistortion}
			\left| \mathbb{E}_c( \mathscr{G}_c )   \right|
			=
			\frac{  \left( 2^{ \frac{p}{2} \pm \mathbf{o}\left( p \right) } \right)^2 - \left( 2^{ \frac{p}{2} \pm \mathbf{o}\left( p \right) }  \right)  }{2} 
			\text{ ,}
		\end{equation} 
		which we know it is possible because of the Borel normality of $ y $ \cite{Calude1994,Calude2002}. 
		Note that, in this case, our construction of the simple MAGs $ \mathscr{G}_c $ ensures that the number of possible composite vertices only varies in accordance with the number of $1$'s in $ y $.
		Therefore, together with basic inequalities in AIT, we have that 
		\begin{equation*}\label{equationInformationdeficiencyinMAGs}
			\begin{aligned}
				\mathbf{K}( \left< \tau( \mathscr{G}_c ) \right> )
				& \leq \mathbf{K}( \left< \mathscr{E}( \mathscr{G}_c  ) ) \right>  + \mathbf{O}(1) \leq \\ 
				& \leq \mathbf{K}( x ) + \mathbf{K}( \left< \mathscr{E}( \mathscr{G}_c  ) \right> \, | \, x  ) + \mathbf{O}(1) \leq \\
				& \leq \mathbf{O}\Big( \log_{2}\big(  \mathbf{K}( \left< \tau( \mathscr{G}_c ) \right> ) \big) \Big) + \mathbf{K}( \left< \mathscr{E}( \mathscr{G}_c  ) \right> \, | \, x  ).
			\end{aligned}
		\end{equation*}
		Finally, the proof of $ \mathbf{K}( \left< \tau( \mathscr{G}_c ) \right> ) + \mathbf{O}(1)
		\geq
		\mathbf{K}( \left< \mathscr{E}( \mathscr{G}_c  ) \right> \, | \, x  )  $ follows directly from Lemma~\ref{lemmaBasicMAGandstrings}. \qed
	\end{proof}
\end{theorem}
The reader is then invited to note that the proof of Theorem~\ref{BthmMAGgivencompaniontuple} also works for many other forms of companion tuples $ \tau( \mathscr{G}_c ) $, as long as Equations~\eqref{equationKrandomrealandcompositeedgesetstring}, \eqref{equationCompaniontuplefromprefixesofrandomreals}, \eqref{equationPrefixesofrandomrealsfromcompaniontuple}, and~\eqref{equationLogarithmiccompressiblechasracteristicstring1} hold.
For example, keep $w$ being a long enough finite initial segment of a $1$-random real number $y$ and then define $ \tau( \mathscr{G}_c ) = \left( | \mathscr{A}( \mathscr{G}_c )[1] |, \dots , | \mathscr{A}( \mathscr{G}_c )[p] | \right) $ such that
\begin{equation*}
	\begin{aligned}
		\mathscr{A}( \mathscr{G}_c )[i] = \left\{ 1 , \dots , \left( f_1(p) + f_2(p) \right) \right\} &\iff \text{the $i$-th digit of $w$ is $1$} \\
		\mathscr{A}( \mathscr{G}_c )[i] = \left\{ 1 , \dots , f_1(p) \right\} &\iff \text{the $i$-th digit of $w$ is $0$}
		\text{ ,} 
	\end{aligned}
\end{equation*}
where $ f_1 : \begin{array}{ccc} \mathbb{N} & \to & \mathbb{N}\setminus\{ 0 \} \end{array} $ and 
$ f_2 : \begin{array}{ccc} \mathbb{N} & \to & \mathbb{Z}\setminus\{ 0 \} \end{array} $ are arbitrary total computable functions.

Moreover, as a consequence of Theorem~\ref{BthmMAGgivencompaniontuple}, we show in Corollary~\ref{corResponseletterreply1} below a phenomenon that can only occur for families of objects embedded into \emph{arbitrarily large} and \emph{non-uniform} multidimensional spaces.
Note that the companion tuple completely determines the discrete \emph{multidimensional space} of the MAGs in which $ \mathscr{A}(\mathscr{G} )[ i ] = \{ 1, \dots , \left| \mathscr{A}(\mathscr{G} )[ i ] \right| \} \subset \mathbb{N} $ holds for every $ i \leq p $.
In the particular case $  \mathscr{A}( \mathscr{G} )[i]  =  \mathscr{A}( \mathscr{G} )[j]  $ holds for every $ i , j \leq p $, we say the multidimensional space of the MAG is \emph{uniform}.
Also note that the number of dimensions of a node-aligned multidimensional network that is mathematically represented by a MAG is given by the value $p$, i.e., the order of the MAG.
Thus, arbitrarily large multidimensional spaces formally refers to arbitrarily large values of $p$.

Specifically, Corollary~\ref{corResponseletterreply1} shows that there are two infinite sets of objects (in particular, one of data representations of multiaspect graphs and the other of strings) whose every member of one set is an encoding of a member of the other, but these members of the two sets are not always equivalent in terms of algorithmic information, which is a phenomenon that some may deem to be counter-intuitive at first glance: 
\begin{corollary}\label{corResponseletterreply1}
	There is an infinite family $F$ of simple MAGs and an infinite set $X$ of the correspondent characteristic strings such that, for every constant $ c \in \mathbb{N} $, there are $ \mathscr{G}_c \in F $ and $ x \in X $, where $x$ is the characteristic string of $ \mathscr{G}_c $ and
	\begin{equation}\label{equationCorollaryResponseletterreply1}
		\mathbf{O}\Big( \log_{2}\big( \mathbf{K}( \left< \mathscr{E}( \mathscr{G}_c  ) \right> ) \big) \Big) > c + \mathbf{K}(x)
		\text{ .}
	\end{equation}
	
	\begin{proof}
		Let $ c \in \mathbb{N} $ be arbitrary.
		Then, in order to construct the family $F$, it suffices to select an infinite number of finite initial segments of a $1$-random infinite binary sequence $y$ such that, for each selected $ y \upharpoonright_{n} $, we choose another $k>n$ with 
		\begin{equation}\label{equationLargerandlargerinitialsegments}
			\mathbf{K}( y \upharpoonright_{k} )
			\geq c + \mathbf{K}( y \upharpoonright_{n} ) + \mathbf{O}\Big( \log_{2}\big(  \mathbf{K}( y \upharpoonright_{k} ) \big) \Big).
		\end{equation}
		This procedure can be applied infinitely many times because $y$ is $1$-random. 
		Now, from the proof of Theorem~\ref{BthmMAGgivencompaniontuple}, construct an infinite set of companion tuples based on these initial segments of $y$.
		From each of these companion tuples, construct the characteristic strings in the same way as in the proof of Theorem~\ref{BthmMAGgivencompaniontuple}.
		Finally, the desired inequality in Equation~\eqref{equationCorollaryResponseletterreply1} then follows from Theorem~\ref{BthmMAGgivencompaniontuple} and Equation~\eqref{equationLargerandlargerinitialsegments}.
		\qed
	\end{proof}
\end{corollary}
We can now combine Corollary~\ref{corResponseletterreply1} with Theorems~\ref{thmMAGisomorphism} and \ref{BthmMAGgivencompaniontuple} in order to show that, although for every MAG there is a graph that is isomorphic to this MAG, they are not always equivalent in terms of algorithmic information, where in fact the distortion may be exponential with respect to the algorithmic information of the graph:
\begin{corollary}\label{corResponseletterreply2}
	There are an infinite family $ F_1 $ of simple MAGs and an infinite family $F_2$ of classical graphs, where every classical graph in $F_2$ is MAG-graph-isomorphic to at least one MAG in $F_1$, such that, for every constant $ c \in \mathbb{N} $, there are $ \mathscr{G}_c \in F_1 $ and a $ G_{ \mathscr{G}_c } \in F_2 $ that is MAG-graph-isomorphic to $ \mathscr{G}_c $, where
	\[
	\mathbf{O}\Big( \log_{2}\big( \mathbf{K}( \left< \mathscr{E}( \mathscr{G}_c  ) \right> ) \big) \Big) > c + \mathbf{K}( \left< E\left( G_{ \mathscr{G}_c } \right) \right> )
	\text{.}
	\]
	
	\begin{proof}
		Let $ F_1 $ be an infinite family of simple MAGs that satisfies Corollary~\ref{corResponseletterreply1}.
		Let $ F_2 $ be a family composed of the classical graphs that are MAG-graph-isomorphic to the MAGs in $ F_1 $.
		Then, for every $ \mathscr{G}_c \in F_1 $ and $ G_{ \mathscr{G}_c } \in F_2 $ that is MAG-graph-isomorphic to $ \mathscr{G}_c $, both have the same characteristic string.
		Remember that, for classical graphs, Equation~\eqref{lemmaBasicMAGandstrings3} holds in the form $ \mathbf{K}( x ) = \mathbf{K}\left( \left< E\left( G_{ \mathscr{G}_c } \right) \right> \right)  \pm \mathbf{O}\left( 1 \right)  $.
		Finally, the proof then follows from Corollary~\ref{corResponseletterreply1}.
		\qed
	\end{proof}
\end{corollary}

\section{Conclusion}\label{sec:conc}

This article presented mathematical results on the limitations for algorithmic information theory (AIT) applied to the study of multidimensional networks with a large number of non-uniform node dimensions (i.e., aspects).
In the case of importing previous approaches for graphs or monoplex networks to node-aligned multidimensional networks,
we demonstrated in Theorem~\ref{BthmMAGgivencompaniontuple}, Corollary~\ref{corResponseletterreply1}, and Corollary~\ref{corResponseletterreply2} the existence of worst-case distortions for network complexity analysis based on network information content or lossless compressibility.
When comparing a logarithmically compressible network topology embedded into a high-algorithmic-complexity multidimensional space with this low-algorithmic-complexity network topology embedded into a unidimensional space, we showed that, in the general case, there are algorithmic complexity distortions that grow linearly with the number of aspects and exponentially with respect to the algorithmic complexity of the monoplex network. 
These distortions occur even though both the multidimensional network and the monoplex network are isomorphic structures.

These results show that a more careful analysis should be taken with purpose of evaluating how the number of distinct aspects, the respective sizes of each aspect, and the ordering that these might be encoded affect the algorithmic information of the whole network.
This way, the present article highlights the importance of: (i)~taking into account the algorithmic complexity of the data structure itself; and (ii)~going beyond the algorithmic complexity of encoding multidimensional networks with characteristic strings or adjacency matrices, for instance. 
Unlike graphs (or monoplex networks), the irreducible information content of a multidimensional network may be highly dependent on the choice of the encoded isomorphic copy.

As we have only dealt with node-aligned multidimensional networks in the form of MAGs, future research is needed for establishing worst-case scenarios when the multidimensional network is not node aligned.



\subsubsection*{Acknowledgments}
{\small Authors acknowledge the partial support from CNPq: F. S. Abrah\~{a}o (301.322/2020-1 ), K. Wehmuth (303.193/2020-4), and A. Ziviani (310.201/2019-5). Authors acknowledge the INCT in Data Science – INCT-CiD (CNPq 465.560/2014-8) and FAPERJ (E-26/203.046/2017). We also thank Cristian Calude, Mikhail Prokopenko, and Gregory Chaitin for suggestions and directions on related topics investigated in this article.}


\bibliographystyle{splncs03}
\bibliography{2.2.1-CompleteRefs-Felipe.bib}

\end{document}